\documentclass{article}

\usepackage[final]{neurips_2023}
\usepackage{verbatim}
\usepackage{amsmath}

\usepackage[utf8]{inputenc} 
\usepackage[T1]{fontenc}    
\usepackage{hyperref}       
\usepackage{url}            
\usepackage{booktabs}       
\usepackage{amsfonts}       
\usepackage{nicefrac}       
\usepackage{microtype}      
\usepackage{xcolor}         
\usepackage{natbib}
\usepackage{siunitx}
\usepackage{graphicx}
\usepackage{caption}
\usepackage[affil-it]{authblk}

\usepackage{arydshln}
\title{A Practical Probabilistic Benchmark for AI Weather Models}

\author[1,*]{Noah D. Brenowitz}
\author[1]{Yair Cohen}
\author[1]{Jaideep Pathak}
\author[2]{Ankur Mahesh}
\author[1]{Boris Bonev}
\author[1]{Thorsten Kurth}
\author[1]{Dale R. Durran}
\author[3]{Peter Harrington}
\author[1]{Michael S. Pritchard}

\affil[1]{NVIDIA, Santa Clara, CA.}
\affil[2]{University of California, Berkeley, CA}
\affil[3]{Lawrence Berkeley National Laboratory, Berkeley, CA}
\affil[*]{Corresponding author: \texttt{nbrenowitz@nvidia.com}}

\begin{document}
\maketitle

\section*{Key Points}
\begin{itemize}
    \item Lagged ensembling is a practical, quantitative, and parameter-free framework for benchmarking weather models
    \item Lagged ensembles of some recent data-driven forecasts are under-dispersive despite claims of ``state of the art'' deterministic skill
    \item Training data-driven models with multi-step loss functions damages ensemble calibration
\end{itemize}

\section*{Abstract}

Since the weather is chaotic, it is necessary to forecast an ensemble of future states. Recently, multiple AI weather models have emerged claiming breakthroughs in deterministic skill. Unfortunately, it is hard to fairly compare ensembles of AI forecasts because variations in ensembling methodology become confounding and the baseline data volume is immense. We address this by scoring lagged initial condition ensembles---whereby an ensemble can be constructed from a library of deterministic hindcasts. This allows the first parameter-free intercomparison of leading AI weather models' probabilistic skill against an operational baseline. Lagged ensembles of the two leading AI weather models, GraphCast and Pangu, perform similarly even though the former outperforms the latter in deterministic scoring. These results are elaborated upon by sensitivity tests showing that commonly used multiple time-step loss functions damage ensemble calibration.

\section*{Plain Language Summary}

2023 was a seminal year for data-driven weather forecasts with several high-profile publications claiming that AI outperformed traditional physics-based approaches to weather forecasts. These claims are mostly supported by scoring deterministic forecasts, even though it is widely known that forecasting is a probabilistic problem.  Probabilistic intercomparisons have proved challenging because of the data volumes involved and because they are confounded by particulars of how probabilistic forecasts are built. As a workaround, we propose benchmarking weather forecasts using lagged ensemble forecasting where forecasts initialized at different times are considered independent samples. When benchmarked in this way, we show that some AI models achieve better deterministic scores by reducing the variance of their forecasts at the cost of physical realism.

\section{Introduction}

Recently data-driven weather models have begun to outperform state-of-the-art physics-based models both in terms of deterministic skill \citep{lam2023learning,bi2023accurate,chen2023fengwu,li2023fuxi} and probabilistic forecasting with large ensembles \citep{price2023gencast,kochkov2023neural,FuXiEns-pb,Mahesh2024-ip}.

Despite these recent successes in probabilistic forecast, benchmarking probabilistic AI models remains a substantial challenge due to the latency of scoring large hind-cast ensembles and the difficulty obtaining baseline data.
On the one hand, large data repositories like WeatherBench 2.0 \citep{rasp2023weatherbench} and TIGGE \citep{TIGGE} have greatly improved access to baseline data though not eliminated the challenge.
However the latency issue is serious since ML model development is an inherently iterative process requiring rapid feedback on the end-to-end performance of the models. It is common to perform multiple evaluations a week or even during a single model training. In many ways (e.g. storage, I/O), a full ensemble scoring is a more substantial computation than the training itself, so by the time one model is scored the rest of the team has moved on.

There is also a need for benchmarking deterministic forecasts from the lens of probabilistic skill.
Even the intrinsically probabilistic models like Gencast \citep{price2023gencast} have been built off strong deterministic backbones like Graphcast \citep{lam2023learning}, so we may expect that improving the realism of individual ensemble members could improve the overall probabilistic forecast.
However, scores like root-mean-squared-error (RMSE) are easy to improve trivially by reducing the variance of a forecast.
By training on mean squared error \citep{bonev2023spherical, lam2023learning} or mean absolute error \citep{bi2023accurate, li2023fuxi} as a key part of the loss function, and correspondingly their performance has often been compared to state-of-the-art  deterministic forecasts \citep{bi2023accurate, bonev2023spherical, li2023fuxi, lam2023learning}. Yet this is unsatisfying as the Root Mean Squared Error (RMSE) for a forecast from a perfect deterministic model saturates at a value $\sqrt{2}$ higher than that for a perfect ensemble forecast as derived theoretically by \citet{leith1974theoretical} and demonstrated for an ensemble of AI weather models by \citet{weyn2021sub}. 
There has been substantial confusion in the literature around this topic.
For example, Graphcast relied on variance reduction via multistep finetuning to improve its scores relative to IFS, but the approach was discarded in the follow-up paper GenCast. 
Any model claiming better deterministic skill at long lead times does so by artificially reducing the variance of the forecast, which will lead it to underestimate the chances of unlikely events.

The purpose of this paper is to present a consistent and easy-to-use method for evaluating deterministic weather forecasts from the lens of probabilistic skill; to show how this clarifies the pitfalls of using only deterministic scoring for model comparison, and to add complementary perspective to ongoing discussions about how best to optimize AI weather predictions. We aim to address the three challenges.

The first is the difficulty of attributing changes in performance to backbone deterministic model rather than the surrounding ensembling framework.
Our motivation to control the ensembling system is similar to \citet{Magnusson2022-xn}, who disentangled the performance of the data assimilation system from model physics by comparing forecasts initialized with the same state.
For ML models the deterministic backbone will typically be a neural network trained on a regression task while for physics it would be either a deterministic GCM or its dynamical core.
Traditional forecast models benefit from the modularity of physics and idealized benchmarks of individual components such as the dynamical core \citep{jablonowski2006baroclinic}, but ML methods can only be tested end-to-end which confounds comparisons given the assortment of frameworks in practice both for representing uncertainty in the initial condition \citep{toth1997ensemble,Molteni1996-hp, leutbecher2005ensemble} and the model itself \citep{palmer2009stochastic}.
As for ML forecast systems, there are similarly a large number of training techniques such as variational auto-encoders \citep{FuXiEns-pb}, de-noising diffusion \citep{price2023gencast}, multi checkpoint ensembles (\citep{Mahesh2024-ip,weyn2021sub}) being developed in tandem with specific ML architectures.
Even with the progress in intrinsically stochastic methods, it seems preferable that the deterministic backbone be as realistic as possible.
So it is hard to know which architecture is the most parameter efficient or has the best performance-compute trade-off.
This is as opposed to other ML domains such as computer vision where both the training technique (e.g loss function) and evaluations are more standardized, and as a result small changes in accuracy can be attributed to the architecture.
Therefore, a parameter-free and quantitative scoring technique specifically targeted at the deterministic backbone is attractive.

The second challenge is that scoring ensemble forecasts over many lead times requires large data volumes that are not easily transferred between various groups that develop models. The amount of data of a modern ensemble of 50-100 members is prohibitive. Assuming we verify at $0.25^\circ$ resolution, with 12-hour sampling up to 14 days, and 600 initial times, then the baseline data needed for scoring for a single ensemble member of one scalar field (e.g. geopotential height at 500 hPa) is 70GB. For 50 members, this becomes TBs, whereas we experience download speeds of 10 MB/s for TIGGE\citep{TIGGE} and 100 MB/s from WeatherBench 2.0 \citep{rasp2023weatherbench}.
Thus, especially with the increasing demand from the growing community experimenting with AI weather prediction, simply obtaining the operational baseline is a challenge.
This download cost can be somewhat mitigated by evaluating at coarser resolution like 1.5 degrees or by scoring with the identical scoring routines, but both of these have their limitations.
This data volume barrier will only grow higher as the resolution of global forecasts increases.

The final challenge is ease and speed of ensemble scoring itself. Once attained, a full operational ensemble baseline becomes a challenging amount of data to use for the purpose of AI model validation. The score that guides AI optimizations should be fast to compute, and ideally in an automated fashion to provide rapid feedback to the ML model developer. Running large ensembles is still slow for high-throughput scoring, even though data-driven models are dramatically faster. 

In this article, we propose a simple method for benchmarking that addresses all
of this challenges, and is suitable for the day-to-day development of improving
deterministic backbone architectures and comparing them with existing
archives of forecast data.
Specifically, we propose a decades-old idea - lagged ensemble forecasts (LEF) - as convenient ensemble for rapid benchmarking.

\citet{Hoffman1983-sy} introduced LEF as an early ensemble forecast technique, which treats all forecasts initialized within some time window of the current time $t$ as separate ensemble members. Theoretically, it exploits the fact that two deterministic forecasts initialized at different times diverge from one another, similar to an initial condition perturbation, taking advantage of the notion that a consistent perturbation of the initial state is one which perturbs the atmospheric state while respecting the correlations in space and between variables that are present in the data; as such, the previous (or next) unperturbed state of the atmosphere can also be viewed as a valid candidate of the perturbed state at the current time. Despite the fact that it has been supplanted by more modern techniques, LEF is still useful in various forecasting problems \citep{trenary2018monthly, vitart2021lagged, chen2013lagged}. Although the number of ensemble members is limited, because they are drawn directly from the data distribution, no new tuning knobs are added.  

Practically, LEF have several properties that make them ideal for validating and intercomparing the deterministic forecast models that form the core of an operational ensemble forecast system.
LEF provides a \textit{parameter free ensemble forecasting system} \citep{brankovic1990extended}.
As such it can be used to fairly compare between different numerical and machine learning weather models.
Lagged ensembles allow one to compute a probabilistic score from deterministic forecast archives, which are not prohibitive in size and can reasonably be moved around the internet or computed on the fly from data-driven models.
The method is simple, and most importantly the same for all models we consider, though there may be some differences in the analyses used to start the deterministic forecasts.
The technique is especially valuable for comparing ML weather models consistently against physical model baselines. 

In the following sections, we apply LEF to intercompare the IFS and leading AI weather models for the first time, to draw several insights. First, we validate the relevance of LEF by showing that, applied to IFS, it can explain most of the temporal variance of the actual operational ensemble (Sec. \ref{sec:lagged-vs-operational}). Then we apply LEF to re-examine the relative skill of two leading AI weather models, to illustrate that a particular AI intervention -- multi-step fine-tuning - that leads to apparent gains in deterministic skill yields no associated gains in probabilistic skill, consistent with the speculation of \citep{lam2023learning} (Sec. \ref{sec:mainbenchmark}). We then confirm this hypothesis by testing the effects of multi-step fine-tuning across a detailed set of ablations with the spherical Fourier neural operator (SFNO) AI weather model \citep{bonev2023spherical} (Sec. \ref{sec:multistep-training}). We then turn to assessing the role of effective resolution  in the context of SFNO architectures where we find that, similarly to fine tuning, systematically decreasing the effective resolution also reduces the spread of the lagged ensemble. Together, these results support our overall thesis -- that LEF provides efficient, practical guidance on probabilistic skill, with potential to accelerate the development of state of the art probabilistic weather forecasting systems.

\section{Lagged Ensemble Benchmark}

\begin{figure}
\centering
\begin{tabular}{ll}
  (a)   & (b) \\
\includegraphics[width=3.5in]{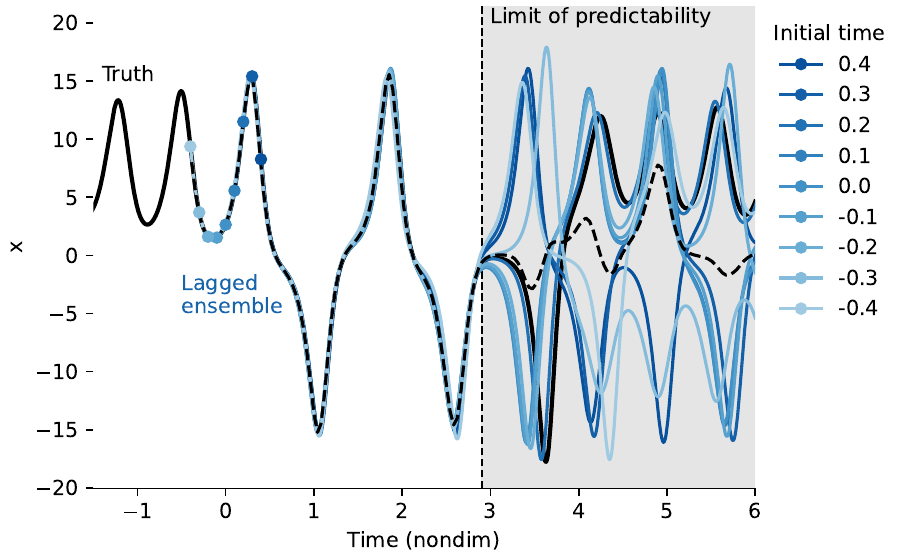}  & \includegraphics[width=2in]{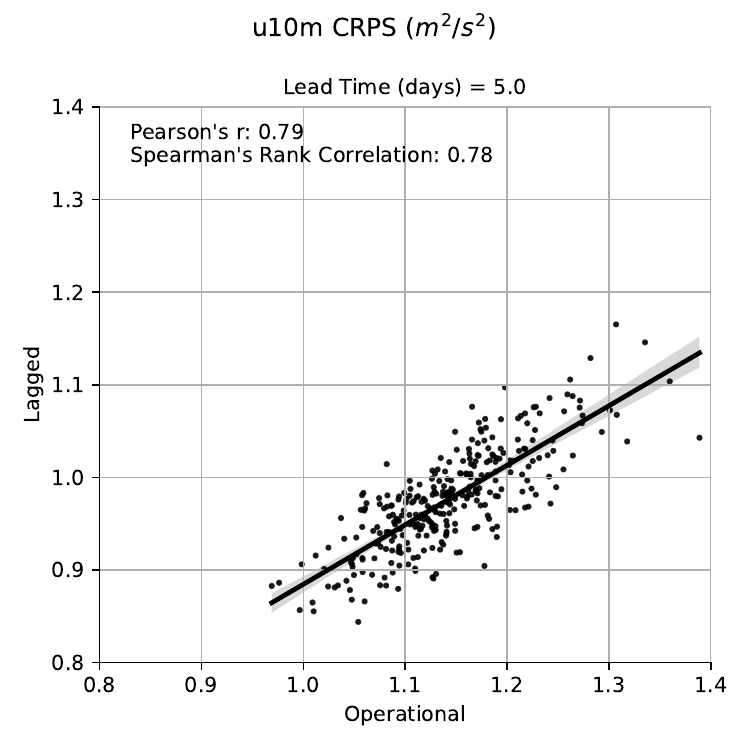}
\end{tabular}
\caption{Overview of lagged ensemble forecasting. (a) A schematic of the method. Each ensemble member (color) is initialized at a different initial time (dots). The true time series (-) and the lagged ensemble average (-~-) are also shown. 
(b) Comparison between the global mean CRPS of \si{500}{hpa} height forecasts from a lagged ensemble and the IFS operational ensemble at a lead time of 5 days. Each dot shows a different initial time. The units of each axes are \unit{m^2/s^2}.\label{fig:schematic}}
\end{figure}

Figure \ref{fig:schematic}a shows a schematic of the lagged ensemble approach on a simple chaotic dynamical system (see Sec \ref{sec:lorenz} for more information). The right-hand side shows how forecasts initialized from different initial times soon diverge from one another, when evaluated at common validation time.
The ensemble mean has a lower variance, but is clearly too smooth and unlike the truth.

To be more precise, let $x(t,t')$ be the state of a simulation valid at time $t$ and initialized at $t'$. The lead time is $t - t'$. Let $h$ be the increment of the forecast interval (often 6 hours in the recent deep-learning weather prediction literature). Then, we develop our scores based on the centered $M$-lagged ensemble given by
\begin{equation}
    x_m(t, t') = x(t, t' - m h),\quad -M \leq m \leq M. \label{eq:ensemble}
\end{equation}
This ensemble has $2M+1$ members and the initial times are centered around $t'$.
While this ensemble cannot be used for a real weather forecast since it requires initial conditions from the future, it is suitable for comparing hindcasts.
Assuming $t$ and $t'$ are uniformly sampled with $h$ simplifies the implementation. 
In practice a lagged ensemble can be generated from any archive of deterministic hindcasts \citep{rasp2023weatherbench,TIGGE}, allowing probabilistic scoring on data volumes far smaller than raw ensemble outputs, which \citet{price2023gencast} pointed out are prohibitive to obtain.

We have implemented a distributed routine for scoring such ensembles using \url{https://github.com/NVIDIA/earth2mip}, an open source software framework.
This tool supports scoring archives on disk, as well as online scoring of data-driven weather models.

The lagged ensemble (and deterministic) forecasts are compared with the corresponding atmospheric states given by ERA5 reanalysis \citep{hersbach2020era5} at $0.25^\circ$ resolution, which is also the training data for all the ML models evaluated in this study.
Hindcasts are made with initial times taken every 12 hours from 2018-01-02 0Z through 2018-10-29 12Z, which is outside the training set for all the models we consider. Like IFS HRES, we consider 10-day forecasts.

There are some important caveats about our choice of ERA5 as verification.
First, to fairly evaluate models with different spatial resolution, we should down-sample to the coarsest available resolution \citep{rasp2023weatherbench}.
Also, ERA5 dataset provides fictitious numerical values on pressure levels that lie below the surface, based on extrapolating a field downwards from the surface in a field-dependent way.
It does not seem fair to evaluate models on how well they predict these unphysical values.
We sidestep these concerns in the present work by only considering models trained with these same data so we leave it to future iterations to addresses these concerns.

We use standard metrics to evaluate the performance of a given lagged ensemble. First, a summary metric of the probability distribution penalizing both over-confidence and inaccuracy is measured using the continuous-ranked-probability score (CRPS)\citep{Gneiting2007-rq}. Second, we evaluate the accuracy of the ensemble mean using the root-mean-square-error (eRMSE, i.e. ensemble-RMSE). Third, we define the deterministic RMSE (dRMSE) as the RMSE of the $m=0$ ensemble member. Finally, we diagnose how well calibrated a lagged ensemble is by comparing the growth of the ensemble standard deviation (spread) to the growth of the eRMSE.
One ensemble is more dispersive than another if its spread is higher for a given level of error (eRMSE).
All metrics are globally averaged and area-weighted.
We use unbiased estimators when applicable for CRPS \citep{Zamo2018-xs}, the ensemble standard deviation (Spread), and spread-error comparisons \citep{Fortin2014-fc}. 
For ensemble mean RMSE, we follow the common practice and report the typical version of the formula.
This is biased in favor of under-dispersive models for small ensembles but no clear alternative is documented in the literature.
Section \ref{sec:metrics} contains more details.

For simplicity we assume all ensemble members are equally weighted. For short lead time forecast, assigning larger weights to more recent lags is preferable as \citet{Hoffman1983-sy} do, but for the purpose of having a parameter-free benchmark, we propose using uniform weights across the ensemble. This somewhat complicates the spread-error and other calibration analyses since the famous 1:1 spread-error relationship assumes ensemble members are interchangeable with each other and the observations \citep{Fortin2014-fc}.
Nonetheless, we do use the spread-error relationship to diagnose if an ensemble is more or less dispersive than another, but will take care to avoid value judgements like ``under'' and ``over'' dispersive.
We will sometimes use the IFS results as a target since the ECWMF operational forecast ensemble built on top of IFS is well-calibrated.
Section \ref{sec:weighted} elaborates how we could extend this analysis to weighted ensembles.

The choice of ensemble size for benchmarking is nuanced.
A small ensemble size limits the usefulness of lagged ensembles for an operational forecast, but for ensemble benchmarking---the focus of our paper---the main concern is that the estimated scores will be (1) noisy and (2) biased. 
To reduce impact of (1) we average over many forecasts and spatial locations. The smoothness of the curves we present suggests that sampling error is not a significant source of uncertainty in our analysis.
As for (2), the use of unbiased estimators in an ensemble with interchangeable members implies that the average scores of an $n$-member ensemble will converge to the same value for $n=2$ or $n=50$.
Since the lagged ensemble members become more interchangeable as the window of initial time narrows, it may be best to choose the smallest $M$ possible.
That said, we have somewhat arbitrarily opted for a 9 member lagged-ensemble ($M=4$).
Figures A1 and A2 show our main conclusions are not sensitive to this choice.
We choose $h=\SI{12}{\hour}$ out of convenience since the HRES data we had on premises was sub-sampled to this time-resolution.
This corresponds to a centered 2-day window around any given initial time with 9 ensemble members.
We only analyze complete ensembles which start at day 2 and end at day 8.

\section{Results}

\subsection{Comparing a lagged and operational ensemble\label{sec:lagged-vs-operational}}

It is natural to wonder to what extent the LEF method can be used as a valid proxy of the operational ensemble. To validate this idea, we compare the lagged ensemble of IFS against the ECMWF operational ensemble \citep{ECMWF-ENS}, see also \ref{m:numerical_model}. By comparing the best deterministic and ensemble forecasting systems from within a single modeling center, we remove many confounding factors in the analysis, although it is worth noting that the IFS and HRES ensembles were generated at different resolutions and with different configurations.

Figure \ref{fig:schematic}b confirms the validity of LEF as a proxy with meaningful operational relevance. At a lead time of 5 days, i.e. sufficiently long that no member of the LEF contains the observations, the CRPS of the lagged and operational ensembles are well-correlated across initial dates: much of the day-to-day variation in global CRPS of the operational ensemble can be predicted from that of the lagged ensemble. As expected, the lagged ensemble is more skillful on average, since it has the artificial advantage that some of its members have shorter lead times than the operational ensemble. We have confirmed that the relationship between LEF CRPS and operational CRPS is similarly monotonic for other variables and lagged ensemble sizes $M$ (cf. Figure A1) and shorter lead times (not shown).

In short, even though there are many differences between these ensemble methods this analysis supports our working hypothesis that LEF is a meaningful probabilistic benchmark of operational relevance. We thus proceed to employ it to perform the first clean probabilistic intercomparison of AI weather models against the IFS LEF baseline.

\subsection{Comparing data-driven and NWP models with the same ensemble method\label{sec:mainbenchmark}}

\begin{figure}
    \centering
    \includegraphics[width=\textwidth]{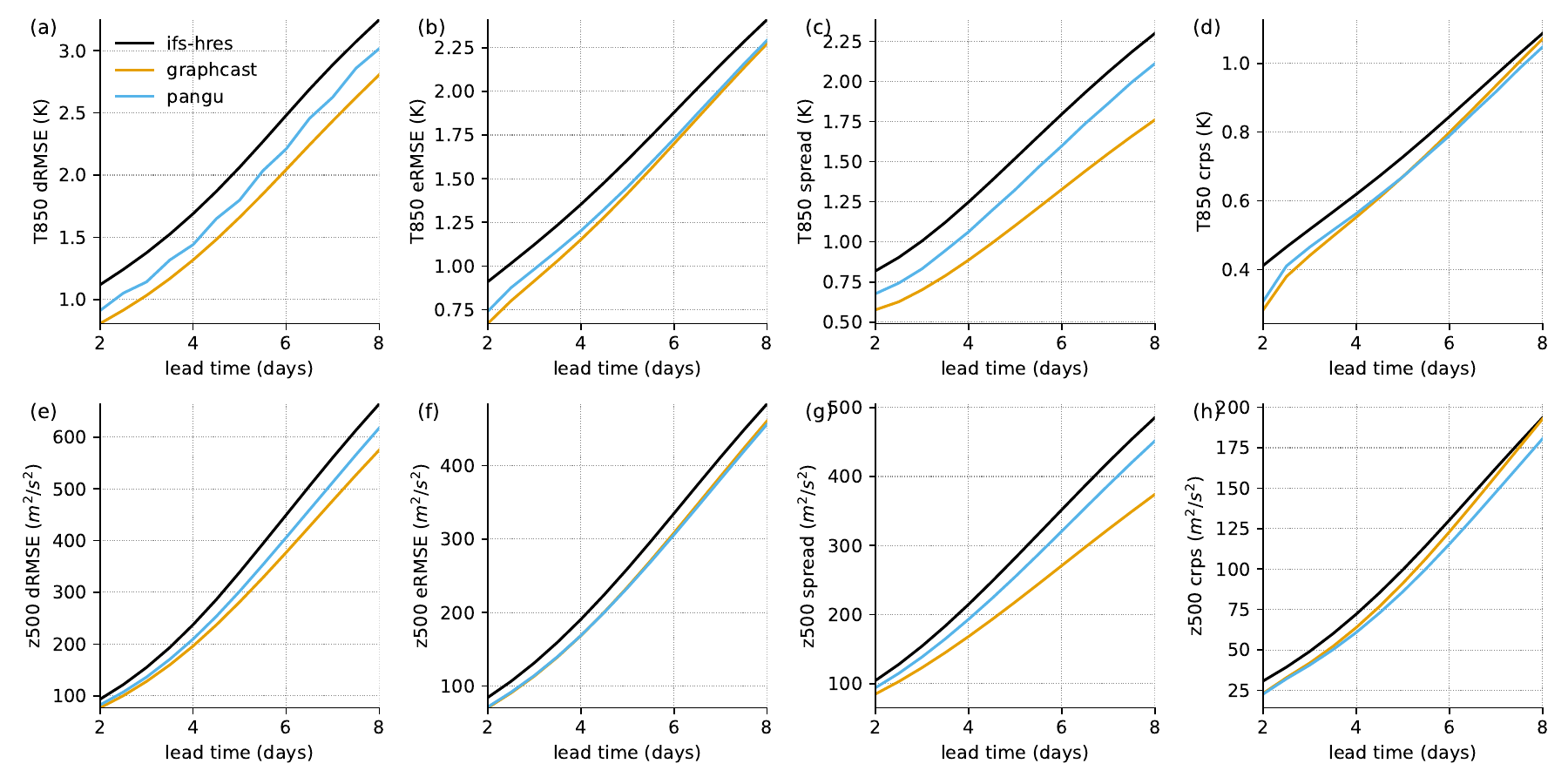} 
    \caption{Z500 and T850 skill comparison for several deterministic forecast systems. From left to right, deterministic RMSE (dRMSE), ensemble RMSE (eRMSE), ensemble spread and CRPS. Ensemble scores are only valid between 2 and 8 days.}
    \label{fig:comparisons}
\end{figure}

Figure \ref{fig:comparisons} compares the ensemble metrics for a set of forecast models including GraphCast, Pangu Weather, and IFS HRES (see Sec. \ref{sec:mlmodels} for details). From the vantage of deterministic skill scores (left column) both of these AI weather models appear to outperform the IFS, and GraphCast especially and increasingly so as a function of lead time. However, the ensemble skill scores -- CRPS and RMSE of the ensemble mean -- tell a different story. All models are virtually tied, subtle differences between the models' ensemble RMSE shrink rather than grow with lead time, and (bottom right) Pangu Weather actually has the lowest Z500 CRPS for many lead times.
The relative performance of Graphcast and Pangu with respect to IFS HRES is similar for different ensemble sizes and channels (see Figure A2).
Overall, these results demonstrate a non-monotonic relationship between deterministic skill and the performance of an ensemble. Also we have shown that some data-driven models, despite being optimized on deterministic loss functions, can outperform physics-based models such as IFS when the same ensemble technique is used.

Note that results for lead times shorter than two days are omitted from our analysis because IFS is initialized with analysis (i.e. knows only about past weather) but we evaluate it against reanalysis (which is informed by future weather). This leads to an initial Z500 error of ~\SI{30}{m}, but for lead times of 2 days and longer this effect is an order of magnitude smaller \citep[Fig S6]{rasp2023weatherbench}.

\subsection{The effect of long-lead time training\label{sec:multistep-training}}

Many of the existing ML weather models employ some method of aggregating loss over time longer than 6h, though the exact methods differ, see Sec. \ref{sec:mlmodels} and Table \ref{tab:ml_models}. Focusing on the two leading models (GraphCast and Pangu), LEF scoring shows that GraphCast is less dispersive than the other models (Figure \ref{fig:comparisons}c,g). That is, for a given level of ensemble mean skill, the ensemble has less variance.

We hypothesize that this is related to the multi-step training protocol and loss function used to train it, wherein successive amounts of auto-regressive time-stepping are unfolded within the optimizer \citep{Brenowitz2018-td,Weyn2020,lam2023learning}, out to lead times of 3 days. Indeed, the minimizer of the MSE loss is known to be the conditional mean, so an ML model trained for long lead times will remove all variance that is not predictable on that time scale.

LEF confirms this hypothesis. We begin by training SFNO models with varying amounts of long-lead time fine-tuning of the autoregressive training. 
The base model is trained for 80 epochs with a context length of a single 6-hour timestep.
We also include results with 2, 3, 5, 9, and 11 steps of autoregressive timestepping in the MSE loss (12h-3 days) to illustrate the effect.
The pre-trained base model is fine-tuned with a smaller learning rate for 20 epochs with a context length of two 6-hour timesteps.
The output from a single forward pass serves as the input for the model to generate a 12-hour forecast. The total accumulated loss from the 6 and 12-hour forecasts is then used as the training objective.
This fine-tuning procedure is sequentially extended, with the context length progressively increasing from 18 hours (3 timesteps) to 3 days (11 timesteps). For each increment in context length, the model is trained for 6 epochs.
NVIDIA has previously released the 11-step model with the name ``FourCastNet v2-small'' \footnote{This model is available on \href{https://catalog.ngc.nvidia.com/orgs/nvidia/teams/modulus/models/modulus_fcnv2_sm}{NVIDIA's NGC Catalog} and also on ECMWF's \href{https://github.com/ecmwf-lab/ai-models-fourcastnetv2}{ai-models platform}.}.

\begin{figure}
    \centering
    \hspace{0mm}\includegraphics[width=0.7\textwidth]{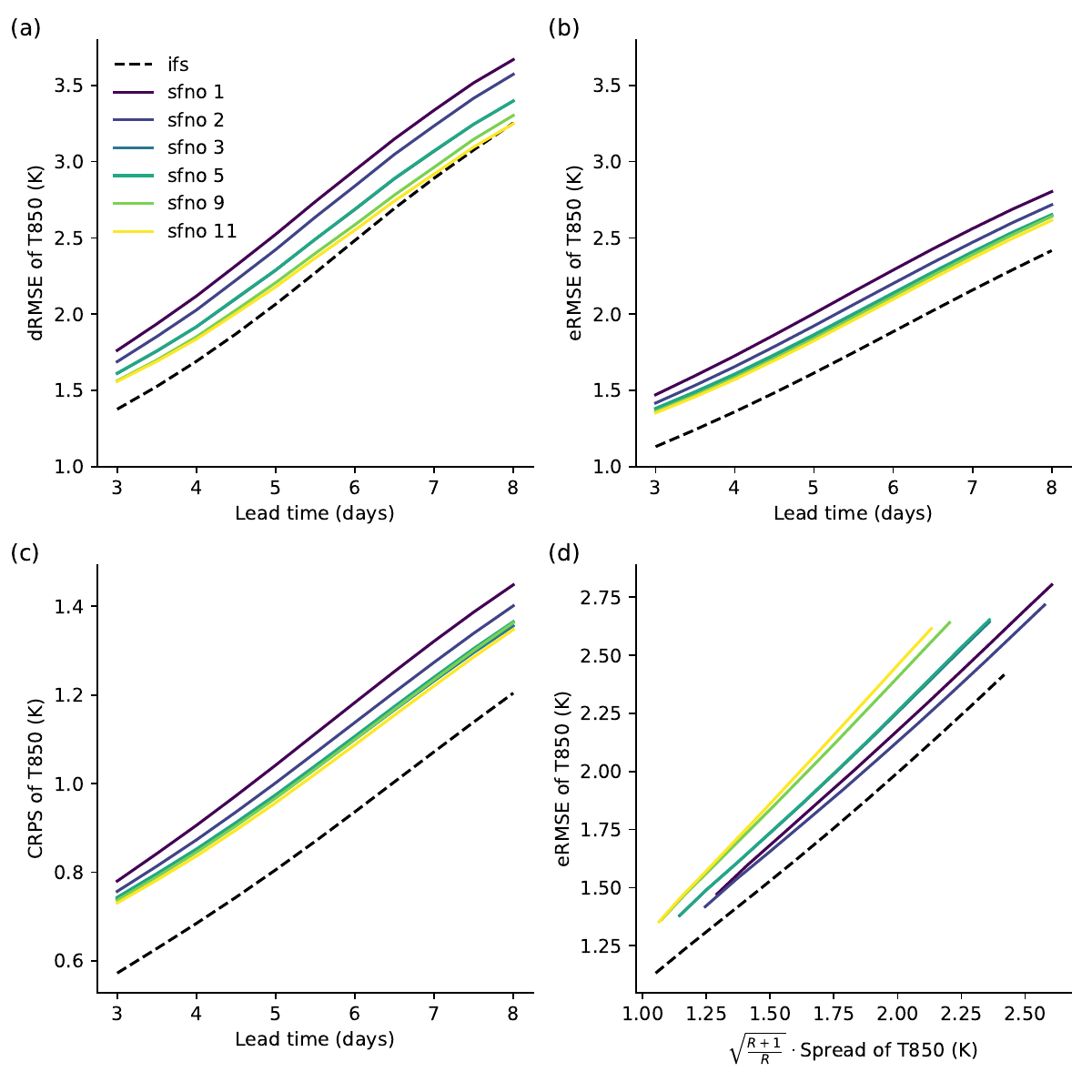} 
    \caption{Using LEF to explore the effect of fine-tuning. Deterministic RMSE (a), ensemble mean RMSE (b) and CRPS (c) as a function of lead time.
    Spread as a function of ensemble mean RMSE (d). The  spread is multiplied by a factor involving the ensemble size $R=2M+1=9$ (see Appendix \ref{sec:metrics}).
    The color lines show a range of fine tuning steps and the dashed line shows the ifs. The field being scored is temperature at 850 hPa (T850). Unlike Figure~\ref{fig:comparisons} the typical biased estimator of CRPS is used (\ref{eq:crps-biased}).}
    \label{fig:fine-tuning}
\end{figure}

Figure \ref{fig:fine-tuning} depicts the results of our LEF analysis applied to this sensitivity test for the 850 hPa temperature field. The deterministic RMSE decreases significantly for models trained with more autoregressive timesteps, even appearing to approach the threshold of IFS deterministic skill limits (panel a).  However, the CRPS and ensemble mean RMSE improve much less than the deterministic RMSE.
The reason is due to plummeting ensemble spread (see Figure~\ref{fig:spread}a-d) where for a given eRMSE the SFNO checkpoints optimized with long-lead time fine-tuning result in monotonic drops in ensemble variance, deviating from the approximately 1:1 error-spread calibration of the LEF IFS ensemble. 

Overall, these results show that autoregressive training behaves similarly to ensemble averaging -- it improves skill by reducing variance. This view is consistent with \citet{lam2023learning} who also noted that long-lead time training came at the expense of a spatially blurred forecast, speculating it optimizes for something in between a deterministic and ensemble mean forecast, a point that has since been elucidated in detail by \citet{price2023gencast} through comparison against more directly probabilistic diffusion-based forecasts. We emphasize that unlike the latter, LEF provides a convenient parameter-free way to trap such issues without the complications of varying approaches to initial condition perturbation.

We also note a relationship between ensemble and time averaging. Since chaotic dynamical systems like the weather forget their initial condition $\lim_{T \rightarrow \infty} E[x(t)|x(t - T)] = E[x(t)]$. This implies that the climate mean $E[X]$ is the best possible forecast for long lead times.
If we further assume ergodicity, one can use time averages of the observed record---known as climatology---to approximate $E[X]$.
To sum up, time averaging, long-lead-time training, and ensemble averaging are roughly equivalent in their ability to reduce the variance of a forecast and thus improve its deterministic skill.

\subsection{Sensitivity to effective resolution\label{sec:scale-factor}}

We now show that it is possible to make a data-driven forecast more dispersive by tuning the architecture.
For computational reasons, many ML models (see Sec. \ref{tab:ml_models}) project the 0.25$^{\circ}$ inputs into a coarse latent space early in the network architecture.
We suspect the resolution of this latent space limits the resolutions the architecture can faithfully generate in autoregressive forecast.

\begin{figure}
    \centering
    \hspace{0mm}\includegraphics[width=\textwidth]{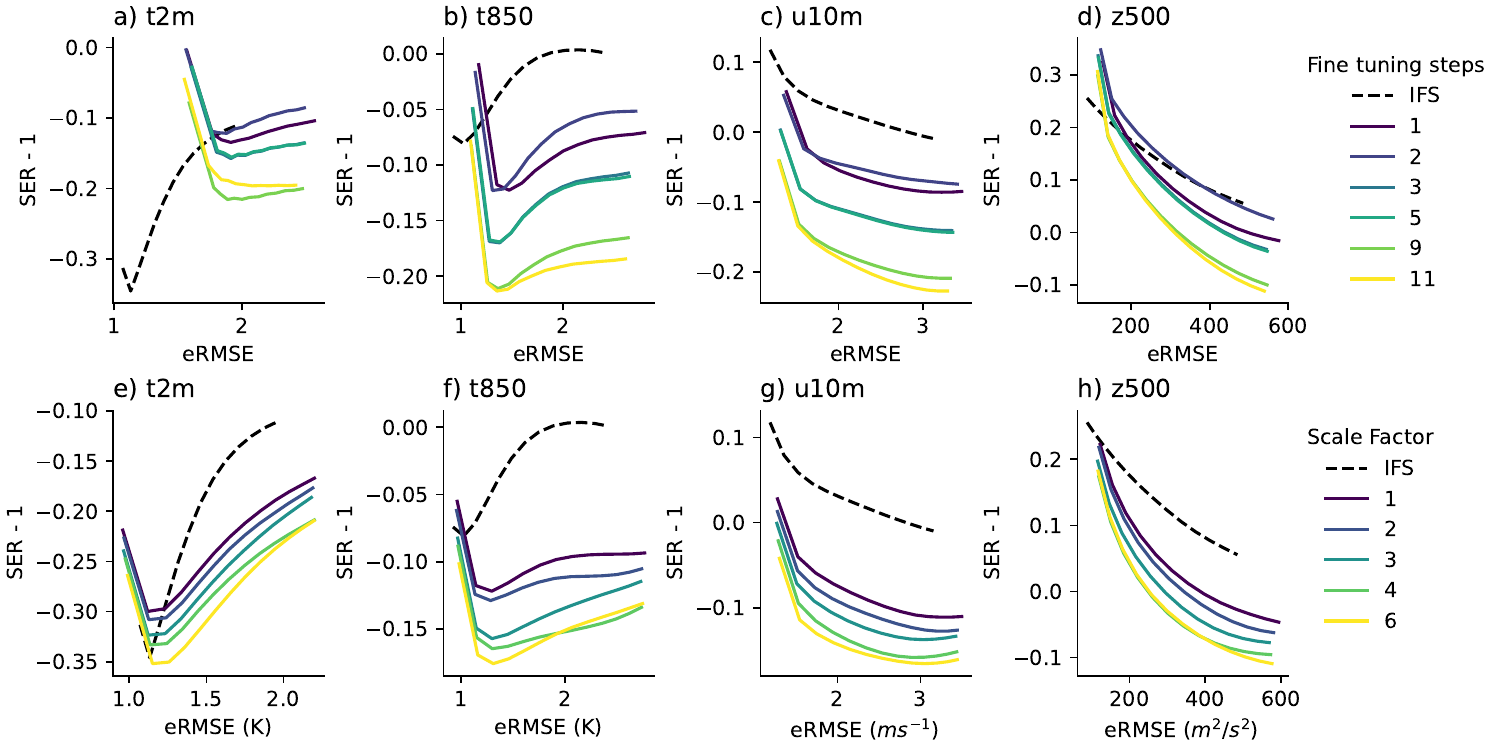}
    \caption{Sensitivity of the ensemble spread to changes in hyperparameters.  (a-d) show results for various fields and differing amounts of auto-regressive fine tuning. (e-h) shows the same, but with a varying scale factor. The bias-corrected spread error ratio is defined by $\text{SER}=\sqrt{\frac{R+1}{R}}\frac{\text{Spread}}{\text{eRMSE}}$.}
    \label{fig:spread}
\end{figure}

We explore this hypothesis by varying the user-specified scale-factor parameter of SFNO.
In the SFNO architecture, the first layer downsamples the input by the scale-factor; larger scale factors correspond to lower spatial resolutions.
We train models with scale factors of 6, 4, 3, 2, and 1.
Each model was trained on 64 NVIDIA A100 GPUs.
SFNOs with scale factor 6, 4, 3, 2, and 1 took 6.9, 7.1, 7.1, 8, and 89 hours to train, respectively. 
Figure~\ref{fig:spread}e-h shows that SFNOs trained with lower scale factors are more dispersive for several fields (higher spread-error ratio).
While these changes did not significantly improve the skill of the model, we suspect that further tuning of the hyper-parameters, especially the embedding dimension, could improve the results.

\section{Discussion and Conclusions}

For 60 years weather modelers have known that we can reduce the RMSE of target variables in a deterministic weather forecast by reducing their variance \citep{leith1974theoretical}.
Yet, the developers of physics-based models typically seek to retain sharpness in single-model forecasts and to increase the variance in ensembles of their model forecasts because those ensembles are typically undispersive. As part of this effort, benchmarks are devised for the dynamical core alone \citep{jablonowski2006baroclinic, thatcher2016moist} to select the numerical methods with the least numerical diffusion, and the resolution is increased in-order to represent physical processes with increasing detail.
Model developers also examine case studies of extreme storms to ensure they produce realistic winds and rainfall \citep{shahrban2016evaluation}.

On the other hand, while there has been substantial progress towards probabilistic forecast \citep{price2023gencast,kochkov2023neural}, much of the data-driven forecast literature has focused primarily on minimizing the error in deterministic forecasts
The easiest way to do this is to reduce the variance of a forecast. 
It is easy to recognize ensemble averaging or space-time filtering as variance-reduction techniques. However, there are many other ML design choices whose effects on ensemble dispersion are less obvious. For instance, we have identified that effective resolution can be a helpful lever on dispersion separately from determinstic skill.

In this work, we devise a simple framework that allows one to compare physics-based and data-driven models with the same ensemble technique.
Within this framework, we highlight that the choice of loss function can produce artificial dissipation akin to ensemble averaging.
Many recent papers have used some form of training involving multi-day lead times including, for example, 
auto-regressive MSE losses\citep{lam2023learning, pathak2022fourcastnet}, replay buffers \citep{chen2023fengwu, li2023fuxi}, or simply predicting very long lead times directly \citep{ClimaX}. Our results show that, when taken too far, such loss functions lead to a poorly calibrated ensemble.

While lagged ensemble scoring is a practical framework for comparing deterministic forecasts, several improvements are possible.
First of all, lagged ensemble members are not interchangeable so we know that a weighted ensemble will outperform our unweighted ensembles here.
The analysis of ensemble calibration is made more ambiguous by this subtlety; we have used the LEF of IFS as a baseline to circumvent this issue.
Finally ensemble size of a lagged ensemble is inherently limited especially if we want the ensemble to be meaningful.
Future work can also focus on applying this technique to more models and working to integrate it into existing community benchmarks which provide a uniform protocol for e.g. output resolution \citep{rasp2023weatherbench}.

Finally, this work presents a convenient way to test a deterministic forecast in isolation from other components.
This is most useful if we assume that the probabilistic forecast systems of the future will be built as they are today: from ensembles of deterministic time steppers.
Indeed AI weather models in this category have been shown to be capable of competitive probabilistic prediction skill \citep{Mahesh2024-ip}.
However, recent work on intrinsically stochastic models suggests an alternative path \citep{price2023gencast,kochkov2023neural}.
While we could score such models with lagged ensembles, this benchmark may give an unfair advantage to deterministic models that have not yet solved the more complex end-to-end probabilistic prediction task.
So final evaluations of end-to-end ensemble forecast systems should still be done with traditional ensemble scoring methods \citep{rasp2023weatherbench}.
Future work could explore more convenient scoring protocols of end-to-end systems---for example with very small ensembles and unbiased scoring rules.

\section*{Acknowledgments}

The computation for this paper was supported in part by the DOE Advanced Scientific Computing Research (ASCR) Leadership Computing Challenge (ALCC) 2023-2024 award `Huge Ensembles of Weather Extremes using the Fourier Forecasting Neural Network' to William Collins (LBNL).  This research was supported by the Director, Office of Science, Office of Biological and Environmental Research of the U.S. Department of Energy under Contract No. DE-AC02-05CH11231 and by the Regional and Global Model Analysis Program area within the Earth and Environmental Systems Modeling Program. The research used resources of the National Energy Research Scientific Computing Center (NERSC), also supported by the Office of Science of the U.S. Department of Energy, under Contract No. DE-AC02-05CH11231.
We further acknowledge Anima Anandkumar and Kamyar Azizzadenesheli for their helpful discussions.

\section*{Open Research}

\paragraph{Software}
A parallel implementation of the lagged ensemble scoring and model inference capability is available at \url{http://github.com/NVIDIA/earth2mip}.

\paragraph{Data}The ERA5 dataset used for training and evaluation is widely available.
We obtained it from a mirror of the Research Data Archive at NCAR \citet{rda} stored at NERSC.
The IFS ensemble and HRES hindcasts that we used as baselines were downloaded from the TIGGE archive at ECWMF.

\paragraph{Model Weights} The graphcast model weights were downloaded from a Google cloud storage bucket \verb|gs://dm_graphcast|.
The PanguWeather weights were downloaded from the ECWMF AI Models repository.
The 11-step trained SFNO checkpoint (cf. Fig \ref{fig:fine-tuning}) is available at \url{https://catalog.ngc.nvidia.com/orgs/nvidia/teams/modulus/models/modulus_fcnv2_sm}.
The earth2mip software automatically downloads these files and can be examined to see the precise URLs used.

\bibliographystyle{plainnat}
\bibliography{reference}

\clearpage
\appendix

\section{Supplemental Information}

\subsection{Lorenz 1963 Dynamical System \label{sec:lorenz}}

\citet{Lorenz1963} introduced a simple differential equation model for chaos in atmospheric convection. The dynamics are given by 
\begin{align*}
\frac{dx}{dt} &= \sigma(y - x), \\
\frac{dy}{dt} &= x(\rho - z) - y, \\
\frac{dz}{dt} &= xy - \beta z.
\end{align*}
In Fig. \ref{fig:schematic} the truth time series is generated using the \verb|scipy.integreate.odeint| routine from SciPy \citep{2020SciPy-NMeth} with the parameters set to $\rho=28$, $\sigma=10.0$, $\beta=8/3$.
The lagged ensemble is generated by forecasts initialized from this true time series without noise perturbation but by setting $\rho=28.14$.
We interpret this as a model error similar to that made by an imperfect weather forecast model.
The trajectories of the lagged ensemble would not diverge without this model error since the dynamics are deterministic and the initial conditions are unperturbed.

\subsection{Numerical models}
\label{m:numerical_model}
The European Centre Medium Range Forecast's (ECMWF) hosts the world's leading global weather forecasting systems. Its deterministic model, the High RESolution forecast (HRES) is a configuration of the Integrated Forecast System (IFS) which produces global 10-day forecasts at about  $0.1^{\circ}$ resolution \citep{ECMWF-IFS}. It is used operationally as the high resolution deterministic member of an ensemble forecasting system (ENS) where additional 50 members at  $0.25^{\circ}$ are used to generate the operational forecast. 

\subsection{Data-driven models\label{sec:mlmodels}}
In the past three years a large number of data driven weather models have been published. Table \ref{tab:ml_models} provides some key information about these models. These models share the training data of ERA5 \citep{hersbach2020era5} and for the most part employ either MAE or MSE loss functions (except FangWu \citep{chen2023fengwu}).
Earlier models use a lower resolution \citep{weyn2019can, rasp2021data, keisler2022forecasting}, while more recent models explore the full ERA5 resolution of $0.25^\circ$.
To save memory and compute, many architectures reduce the resolution of this data internally by embedding it into a coarser latent space.
A common practice is to coarsen the ERA5 input data by a factor (typically 4) while predicting at full resolution (see the "scale factor") column in Table \ref{tab:ml_models}.
The details (and terminology) of such coarsening depends on the architecture, e.g. a spherical harmonics based model like SFNO uses a scale factor, effectively limiting the maximum wave number, GraphCast uses the embedded spatial resolution on its icosahedral grid while a ViT based model like PanguWeather uses patch embedding.  

Many models are trained with autoregressive loss accumulation of different kinds (see Table \ref{tab:ml_models}).
\citet{Brenowitz2018-td} introduced this idea to stabilize machine learning parameterizations, and then \citet{weyn2021sub} applied it to global weather prediction.
Since then, severals works have also used autoregressive loss, including SFNO \citep{bonev2023spherical} and GraphCast \citep{lam2023learning}.
The latter reported a comparison of the effect of number of steps on the RMSE (see figure S34 there) and also note that auto-regressive training encourages GraphCast to blur the forecast.
PanguWeather \citep{bi2023accurate} took a different approach by training several models at different time-steps (namely 1,3,6,24 hours) and interleaving them in the forecast. FuXi \citep{li2023fuxi} combines both methods in the so-called cascade approach where separate models are trained and fine-tuned for different parts of the medium range (0-14 days lead time).
FengWu \citep{chen2023fengwu} used a ``replay buffer'' in which the model is trained to predict the weather at time $t$ from both target at time $t-1$ and from stored model forecasts at time $t-1$; effectively learning to veer back to the target data when provided with initial condition with a lead time error accumulated over several steps. This in practice could encourage the model to learn an approximation of the ensemble mean as well.

Furthermore, most models are evaluated with deterministic metrics (RMSE and ACC), exceptions are reported \citep{bi2023accurate, chen2023fengwu}. 

\begin{table}[h!]
  \centering
  \caption{A list of the most relevant global ML weather models. Here NLL used by \citet{chen2023fengwu} is a negative log-likelihood loss.}
  \label{tab:ml_models}
    \begin{minipage}{\textwidth}
      \centering
      \resizebox{\textwidth}{!}{%
        \begin{tabular}{>{\raggedright\arraybackslash}p{2.0cm} | p{2.0cm} | p{2.0cm} | p{2.0cm} | p{2.0cm} | p{2.0cm} | p{2.0cm} | p{2.0cm}}
          \toprule
          \textbf{Feature} & \textbf{DLWP} & \textbf{FCN-SFNO} & \textbf{GraphCast} & \textbf{Pangu} & \textbf{FuXi} & \textbf{Keisler} & \textbf{FengWu} \\
          \midrule
          Architecture & CNN & SFNO & GNN & swin & Transformer & Enc-Dec  & Transformer \\
          \hdashline
          Parameters & $0.7 \times 10^6$ & $829 \times 10^6$ & $36 \times 10^6$ & $64 \times 10^6$ & -- & $6.7 \times 10^6$ & -- \\
          \hdashline
          Resolution & $2^\circ$ & $0.25^\circ$ & $0.25^\circ$ & $0.25^\circ$ & $0.25^\circ$ & $1^\circ$ & $0.25^\circ$ \\
          \hdashline
          Loss & MSE & MSE & MSE & MAE & MAE & MSE & NLL \\
          \hdashline
          Autoregressive training & 4 $\times$ 6h steps & 12 $\times$ 6h steps & 12 $\times$ 6h steps & interleaving 1,3,6,24h & 12 $\times$ 6h steps + interleaving & 10 $\times$ 6h steps & Replay Buffer \\
          \hdashline
          Scale factor & 4 & 6 & 4 & 4 & -- & -- & -- \\
          \hdashline
          Metrics & RMSE, ACC & RMSE, ACC & RMSE, ACC & RMSE, ACC, CRPS, STD/RMSE & RMSE, ACC, CRPS, STD, SSR & RMSE & ACC, RMSE \\
          \bottomrule
        \end{tabular}
      }
    \end{minipage}
\end{table}

\paragraph{Spherical Fourier Neural Operator (SFNO)}

\citet{bonev2023spherical} introduced a spherical Neural Operator, which respects the underlying geometry of the sphere and its inherent symmetries. The resulting architecture has remarkable stability properties with reported autoregressive rollouts of up to a full year \citep{bonev2023spherical}.

Fourier Neural Operators are closely related to convolutional networks, where convolutions are computed in the continuous domain via the spectral theorem. This makes Neural Operators fundamentally agnostic of the underlying discretisation of input and output signals and enables them to be trained and evaluated on any discretisation that permits the computation of the operations.

In the following, we utilize a range of SFNO models, which differ in model resolution via a scale factor. This scale factor effectively controls the downsampling of the signal and the resolution of the internal representation. A scale factor of one corresponds to no downsampling, whereas a scale factor of two corresponds to a 2x downsampling in the internal representation. Higher downsampling ratios therefore lead to lower effective resolutions and a higher smoothing of the predicted fields.

\paragraph{Pangu Weather}

Pangu Weather \citep{bi2023accurate} introduced a hierarchy of vision transformers trained to take different time steps.
The publicly available models include
\begin{itemize}
    \item 3 hour time step,
    \item 6 hour time step, and
    \item 24-6 interleaved time step.
\end{itemize}

The interleaved time stepper is effectively a 24-hour model, with 6-hour steps used to interleave in between.
For example, to make a 72 hour forecast, requires 3 24 hour steps and no 6 hour steps. A 36 hour forecast requires a single 24 time step and two 6 hour steps. 
We use the same interleaving strategy for each member of the lagged ensemble, so that a -1 lagged member at time $t$ is the same as the 0-lagged member at time $t-1$.
Each lagged member is simply a deterministic forecast at the lagged initial time.



\subsection{Ensemble Verification Metrics\label{sec:metrics}}

Let $\{x_m(t)\}_{m=1,\ldots,R}$ be an ensemble of forecasts of a quantity $y(t)\in \mathbb{R}$. Let $\langle \cdot \rangle $ be a potentially weighted average over many samples $t$, which could be taken over space and time.
We identify $m=1$ with the unperturbed control forecast---in the case of lagged ensembles this is the non-lagged initialization $m=0$ in (\ref{eq:ensemble}).
We present area-weighted global averages of this metric, where the area weighting is given by $\cos(lat)$.
Like \citet{rasp2023weatherbench}, for RMSE and spread, initial time averages are taken across the lagged ensemble at fixed lead time square root, so that the square of RMSE is a time and space-averaged MSE. 

The deterministic RMSE is given by
\begin{equation}
    \text{dRMSE} = \sqrt{\langle|x_0 - y|^2\rangle}.
\end{equation}

When evaluating with a small ensemble size some care is required in defining and interpreting probabilistic metrics. Let the ensemble mean $\overline{x} = R^{-1} \sum_{m=1}^R x_m$.
We use the common definition of ensemble mean MSE:
\begin{equation}
    \text{eRMSE} = \sqrt{\left\langle \left| \overline{x} - y\right|^2 \right\rangle }
\end{equation}
This formula is biased for finite ensembles.
\citet{Zamo2018-xs} give an unbiased estimator of CRPS as
\begin{equation}
    \text{CRPS} = \left\langle \frac{1}{R}\sum_{m=1}^R \left|x_m - y\right| - \frac{1}{2 R (R-1)}\sum_{m,m'=1}^R \left|x_m - x_m'\right| \right\rangle.
\end{equation}
The more commonly used, yet biaesd formula---derived from estimating using the substituting the empirical CDF into the continuous formula---is given by
\begin{equation}
    \text{CRPS}_{biased} = \left\langle \frac{1}{R}\sum_{m=1}^R \left|x_m - y\right| - \frac{1}{2 R^2}\sum_{m,m'=1}^R \left|x_m - x_m'\right| \right\rangle. \label{eq:crps-biased}
\end{equation}
Note the different denominator in the second term leads to $\text{CRPS}_{biased}$ favoring models with lower spread.

The ensemble spread is defined as the unbiased sample standard deviation of the ensemble, which is given by
\begin{equation}
    \text{Spread} = \sqrt{\left\langle \frac{1}{R - 1}\sum_{m=1}^R \left|x_m - \overline{x} \right|^2 \right\rangle}.
\end{equation}
\citet{Fortin2014-fc} show that that a necessary condition for a well calibrated is that a de-biased spread-error ratio (SER) is 1. That is
\begin{equation}
    \text{SER} := \sqrt{\frac{R+1}{R}}\frac{\text{Spread}}{\text{eRMSE}} = 1. \label{eq:ser}
\end{equation}

\subsection{Spread-error correspondence for a weighted ensemble\label{sec:weighted}}

\citep{Fortin2014-fc} derived the famous 1-1 spread-error relationship expected assuming the ensemble members and observations are interchangeable. This is true for initial conditions ensembles, but the members of a lagged ensemble are not interchangeable since the more recent members are more skillful.

Bayesian model averaging (BMA) provides a more general formalism for such cases \citep{Raftery2005-md}. Specifically, let $f_m(t_j)$ be the forecast valid at time $t_j$ but initialized with a lag $m$. The initial time of this forecast is $t_j - h(\ell + m)$ where $\ell h$ is the lead time of the forecast. Let $S_{j} = \bigcup_{m} \{f_m(t_j)\}$  be the set of such forecasts. BMA assumes the distribution of a variable $Y$ drawn from this ensemble is given by
\begin{equation}
p(y| S_j) = \sum_m w_m p(y|f_m),\label{eq:mixture}
\end{equation}
where $w_m$ is the weight of the ensemble member $m$, which can be interpreted as a Bayesian posterior probability that $f_m$ is the correct ensemble member.

As an example, the ``tempering'' approach of \citet{Hoffman1983-sy} can be cast into the BMA framework by letting $p(y|f_m)=\delta(y-f_m)$ where $\delta$ is Dirac's delta function, and the $w_m$ are fit using least squares.
Under this model, the mean and variance of this distribution are
\[
E(Y|S_j) = \sum_m w_m f_m
\]
and
\begin{equation}
Var(Y|S_j)= E((Y - E[Y|S_j])^2) = \sum_m w_m \left( f_m - \sum_m w_m f_m \right)^2.\label{eq:spread-error}
\end{equation}
We can check the consistency of this relationship empirically by further averaging this equation in space and time, and comparing the left-hand-side (error) to the right (spread).
While we do not use this derivation in our results it could be useful for future work.

\subsection{Sensitivity to Ensemble Size}

\begin{figure}[h]
    \centering
    \begin{tabular}{cccc}
M=1 & M=2 & M=3 & M=4 \\
\includegraphics[width=0.25\textwidth]{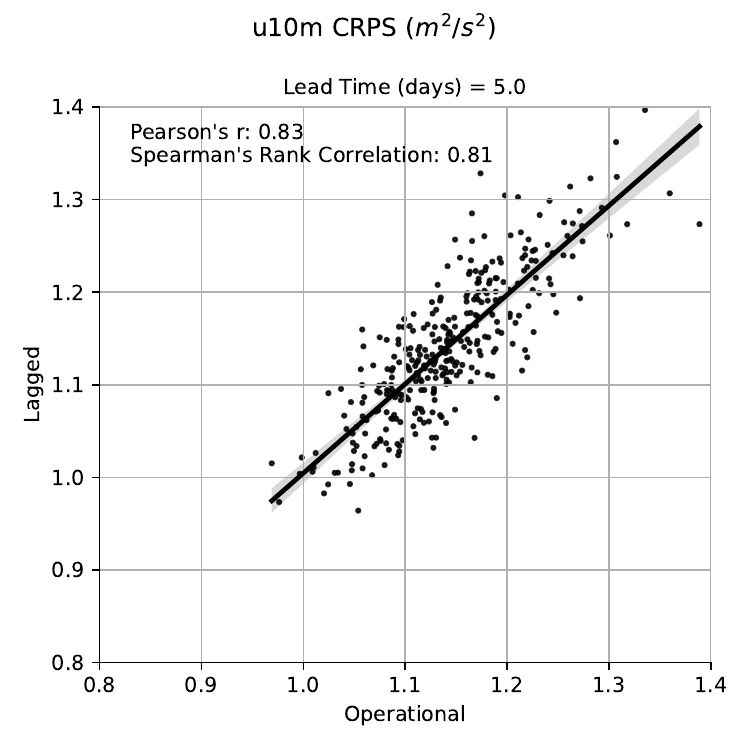}
&
\includegraphics[width=0.25\textwidth]{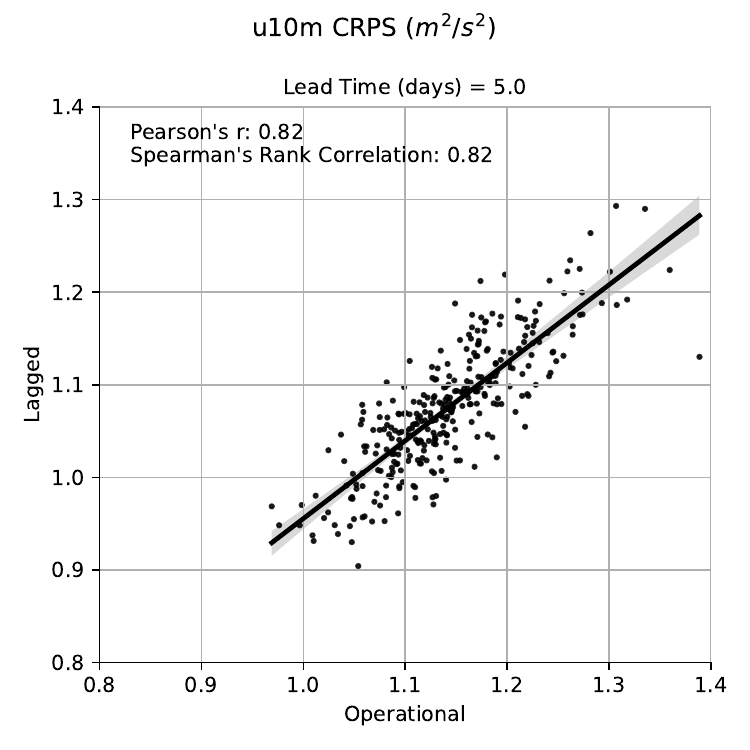}
&
\includegraphics[width=0.25\textwidth]{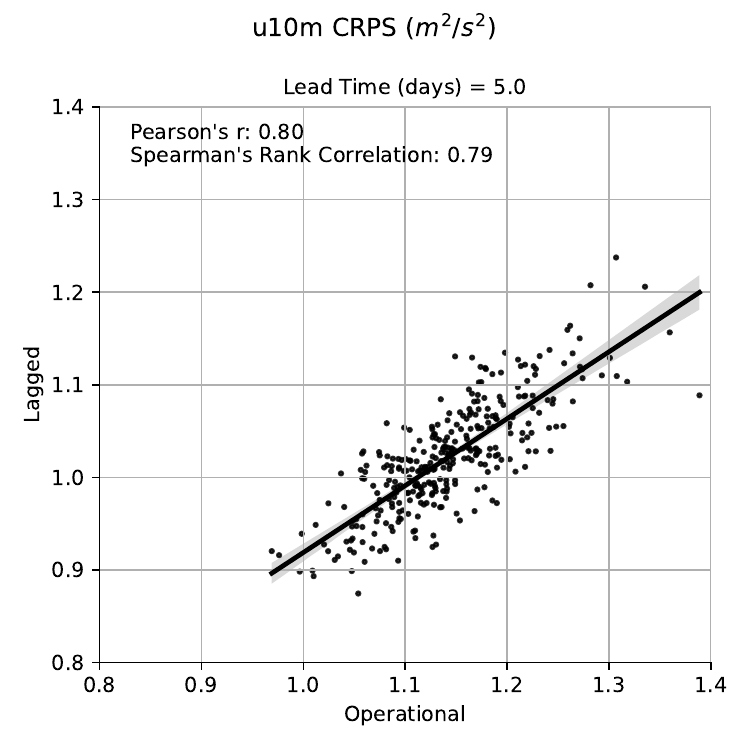}
&
\includegraphics[width=0.25\textwidth]{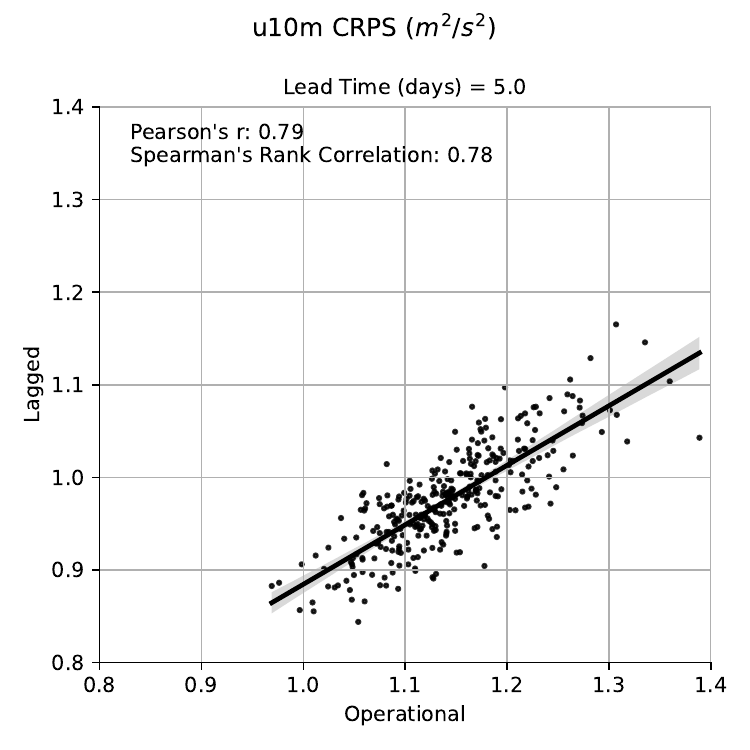}
\\
\includegraphics[width=0.25\textwidth]{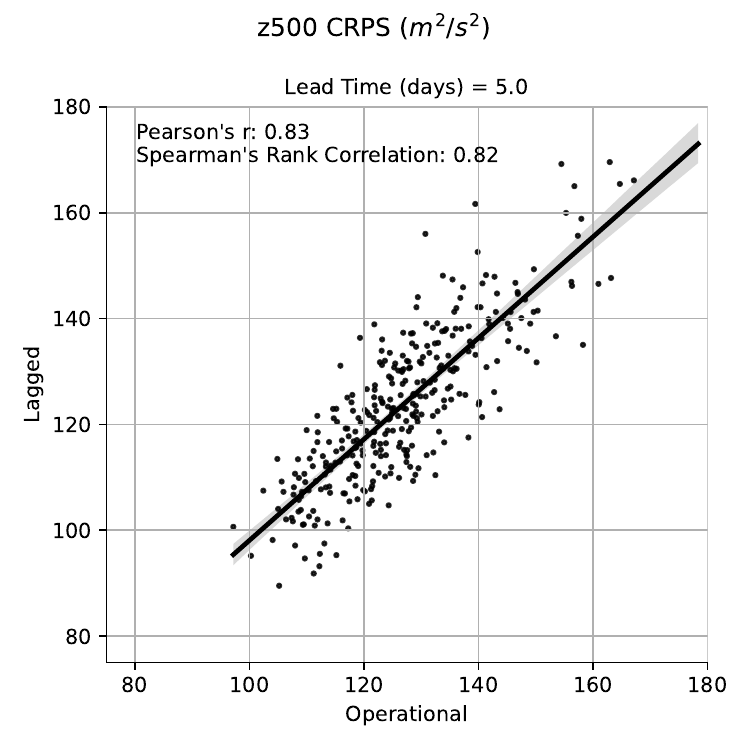}
&
\includegraphics[width=0.25\textwidth]{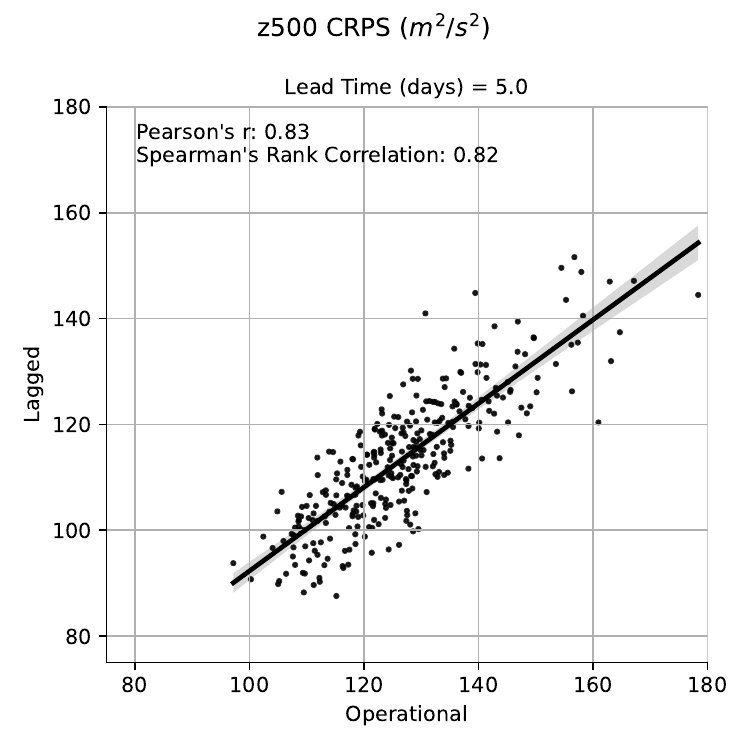}
&
\includegraphics[width=0.25\textwidth]{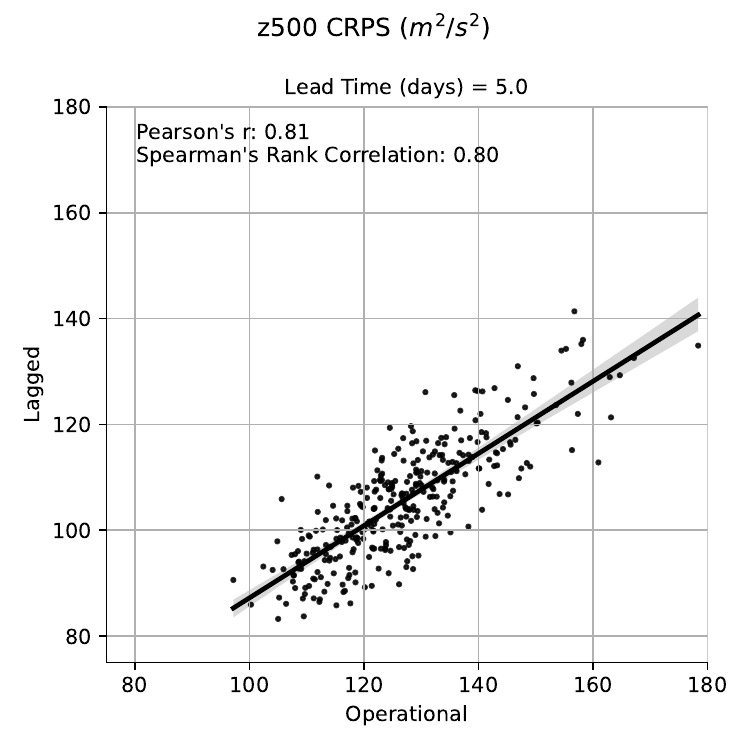}
&
\includegraphics[width=0.25\textwidth]{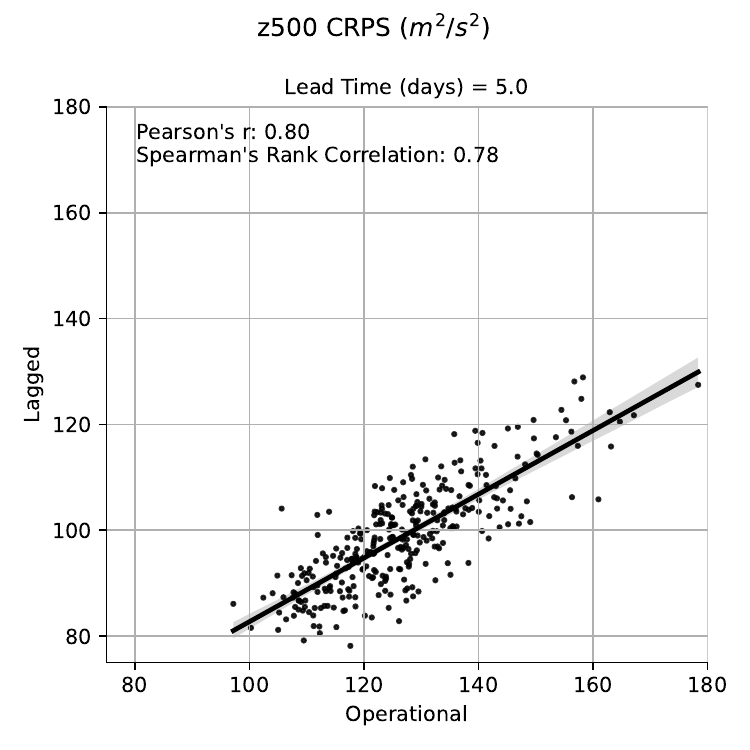}
\end{tabular}

    \caption*{Figure A1. Same as \ref{fig:schematic}b but for differing numbers of lags $M=1,2,3,4$ and for two channels, u10m and z500. The correlation between the lagged and CRPS is similar in all cases. The lagged CRPS has similar quantitative values to the ENS CRPS for $M=1$}
    \label{fig:scatter-sensitivity}
\end{figure}

\begin{figure}[h]
    \centering
    \includegraphics[width=\textwidth]{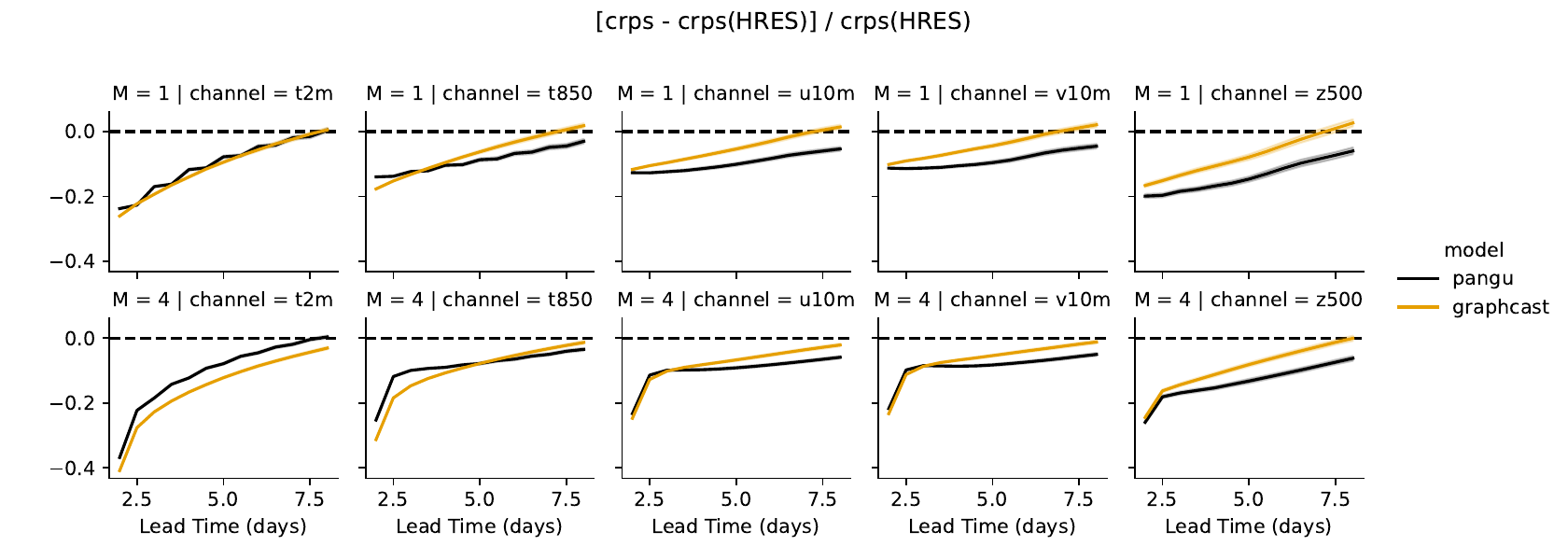}
    \caption*{Figure A2. Sensitivity of relative performance to the lagged ensemble size $M=1,4$. Shows the fractional improvement in CRPS relative to HRES for PanguWeather and Graphcast for several channels.}
    \label{fig:relative-performance-sensitivity}
\end{figure}

\end{document}